\documentclass[%
 reprint,
 amsmath,amssymb,
 aps,
pra,
]{revtex4-1}

\usepackage{graphicx}
\usepackage{dcolumn}
\usepackage{bm}

\begin{document}

\preprint{APS/123-QED}

\title{Flexible quantum private queries based on quantum key distribution}

\author{Fei Gao}
 \altaffiliation[Also at ]{State Key Laboratory of Integrated Service Networks, Xidian University, Xi'an 710071, China}
 \email{gaofei\_bupt@hotmail.com}
\author{Bin Liu}
 \email{lbhitmanbl@gmail.com}
\author{Qiao-Yan Wen}
\affiliation{
 State Key Laboratory of Networking and Switching Technology, Beijing University of Posts and Telecommunications, Beijing, 100876, China
}
\author{Hui Chen}
\affiliation{
 Science and Technology on Communication Security Laboratory, P.O. Box 810, Chengdu, 610041, China
}

\date{\today}

\begin{abstract}
We present a flexible quantum-key-distribution-based protocol for quantum private queries. Similar to M. Jakobi et al's protocol [Phys. Rev. A \textbf{83}, 022301 (2011)], it is loss tolerant, practical and robust against quantum memory attack. Furthermore, our protocol is more flexible and controllable. We show that, by adjusting the value of $\theta$, the average number of the key bits Alice obtains can be located on any fixed value the users wanted for any database size. And the parameter $k$ is generally smaller (even $k=1$ can be achieved) when $\theta<\pi/4$, which implies lower complexity of both quantum and classical communications. Furthermore, the users can choose a smaller $\theta$ to get better database security, or a larger $\theta$ to obtain a lower probability with which Bob can correctly guess the address of Alice's query.
\begin{description}
\item[PACS numbers]
03.67.Dd, 03.67.Hk
\end{description}
\end{abstract}

\pacs{Valid PACS appear here}
\maketitle


\section{\label{sec:level1}Introduction}

Cryptography is the approach to protect data secrecy in public environment. As we know, the security of most classical cryptosystems is based on the assumption of computational complexity and might be susceptible to the strong ability of quantum computation \cite{Shor,Grover}. Fortunately, this difficulty can be overcome by quantum cryptography \cite{BB84,GRTZ02}, where the security is assured by physical principles. With the advantage of higher security, quantum cryptography has attracted a great deal of attention now.

In some cryptographic communications, we need not only protect the security of the transmitted message against eavesdropping from an outside adversary, but also the communicators' individual privacy against each other. Private information retrieval (PIR) \cite{PIR} and symmetrically private information retrieval (SPIR) \cite{SPIR} are protocols for such circumstance. Both of them deal with the problem of private user queries to a database, where Alice wants to obtain one item secretly from Bob's database. Generally it is assumed that Alice knows the address of this item in the database and, for simplicity, the content of it is just a bit. In a PIR protocol the aim is that Alice gets the item she wanted correctly and, at the same time, Bob does not know which item Alice has obtained. SPIR protects not only Alice's privacy but also the security of Bob's database, that is, Alice cannot get other items except the one she wanted in the database. However, the task of SPIR cannot be implemented ideally \cite{Lo1997}. More practically, it is generally required that Alice can elicit sufficiently little content of the database, or at least cannot get the whole database by dishonest operations.

Quantum Private Queries (QPQ) is the quantum scheme for SPIR problem. In 2008 V. Giovannetti et al proposed the first QPQ protocol (GLM protocol) \cite{QPQ08}, where the database is represented by a unitary operation (i.e. oracle operation) and it is performed on the coming query states. In this protocol two query states are needed. One is for getting the wanted information from the database and the other, a superposition state, for checking potential attack from Bob. GLM protocol is cheat sensitive for Alice's privacy: if Bob tries to obtain information on the query he would be discovered by Alice with a certain probability. It ensures perfect data privacy of the database: Alice can obtain at most two items if she perform dishonest queries. Compared with previous schemes GLM protocol displays an exponential reduction in both communication complexity and running-time computational complexity. Furthermore, the security of GLM protocol was deeply analyzed \cite{QPQproof} and a proof-of-principle experimental realization of it was implemented \cite{QPQexp}.

Recently L. Olejnik presented a quantum PIR protocol (O-protocol) \cite{QPIR11}, which is similar to GLM protocol in terms of form. In this protocol the oracle operation and the coding method are subtly selected so that one query state can achieve two aims simultaneously, i.e. obtaining the expected information and checking Bob's potential attack. Here Alice's checking is not a real-time one (like that in GLM protocol) because only one query is executed, but Bob's answer may be wrong if he tries to obtain Alice's privacy, which will leave the trace of his attack and be discovered by Alice later. Moreover, O-protocol can reduce communication complexity further.

Though the above two protocols exhibit significant advantages in theory, they are difficult to implement because when large database is concerned the dimension of the oracle operation will be very high. To solve this problem, M. Jakobi et al gave a new QPQ protocol (J-protocol) based on quantum key distribution (QKD) \cite{QPQ11}, which is the first practical QPQ protocol and quite different from the previous ones. In this protocol SARG04 QKD protocol \cite{SARG04} is utilized to distribute asymmetric key between Alice and Bob, and the whole database is encrypted by the key. Because Alice only knows some bits of the key, she can obtain limited items in the database. Compared with GLM protocol and O-protocol, J-protocol can be easily generalized to large database and it is loss tolerant. Furthermore, it is difficult for Bob to obtain the address of Alice's query even though he tries his best to attack. In this sense, Alice's privacy is protected better in J-protocol. Therefore, this new model of QPQ, i.e. QKD-based QPQ, is very attractive and will become a studying point in the future.

In this paper, inspired by the work of M. Jakobi et al \cite{QPQ11}, we propose a new QPQ protocol based on QKD, which can be seen as a generalization of J-protocol. We show that our protocol preserves all the features of J-protocol and is more flexible. In particular, by adjusting the parameters the expected value of the key bits Alice obtains can be located on a certain number for databases of any size. And it also displays the advantages of lower communication complexity and higher security degree when suitable parameters are selected.

The rest of this paper is organized as follows. In Sec. II we describe our protocol in detail and show its features especially on the flexibility. The security of our protocol is analyzed in Sec. III and Sec.IV is our conclusion.

\section{QPQ protocol based on QKD}

Without loss of generality, suppose there are $N$ items in Bob's database, and Alice has bought one of them and wants to obtain it secretly. The two users can execute the following protocol.

\subsection{The protocol}

(1) Bob prepares a long sequence of photons which are randomly in one of the states $\{|0\rangle,|1\rangle,|0'\rangle,|1'\rangle\}$, and sends them to Alice. Here
\begin{eqnarray}
|0'\rangle=\cos\theta|0\rangle+\sin\theta|1\rangle, \nonumber\\
|1'\rangle=\sin\theta|0\rangle-\cos\theta|1\rangle
\label{eq:one},
\end{eqnarray}
and $|0\rangle$ and $|1\rangle$ represent bit 0, while $|0'\rangle$ and $|1'\rangle$ code for bit 1. The parameter $\theta\in(0,\pi/2)$ can be selected continuously according to particular situations, which will be demonstrated below.

(2) Alice randomly measures each received photon in the basis $B=\{|0\rangle,|1\rangle\}$ or the basis $B'=\{|0'\rangle,|1'\rangle\}$. Obviously this measurement does not allow her to infer the value of the bit sent by Bob.

(3) Alice announces in which instances she has successfully detected the qubit. The bits carried by the lost photons are disregarded. Note that Alice cannot cheat by lying in this step (e.g. announcing a photon lost when she gets an unwanted measurement result). This is because till now Alice has no information about the sent bits and she cannot obtain any benefit by such a lie. Therefore, this protocol is completely loss tolerant, which is similar to that in J-protocol.

(4) For each qubit that Alice has successfully measured, Bob announces one bit 0 or 1, where 0 represents this qubit is originally in the state $|0\rangle$ or $|0'\rangle$, while 1 implies the qubit is $|1\rangle$ or $|1'\rangle$.

(5) Alice interprets her measurement results in step 2. According to her measurement result and Bob's declaration, Alice can obtain the sent bit with a certain probability. This process is similar to that in B92 QKD protocol \cite{B92}. For example, if Bob's declaration is 0 and Alice's measurement result is $|1\rangle$ ($|1'\rangle$), she knows the qubit must be in the state $|0'\rangle$ ($|0\rangle$) before the measurement and then the sent bit is 1 (0). As a result, with Bob's declaration in step 4, Alice's measurements will yield $p=\sin^2\theta/2$ of conclusive results and $1-p$ of inconclusive ones. Both conclusive and inconclusive results are stored. by this way Alice and Bob now share a raw key $K^r$ which is known completely to Bob and partly to Alice (she knows $p=\sin^2\theta/2$ of the whole).

(6) Two users execute postprocessing to the key so that Alice's known bits in the key are reduced to 1 bit or a little more. Without loss of generality, suppose the length of the raw key shared between Alice and Bob is $kN$. Here the natural number $k$ is a parameter and we will discuss its value later. Alice and Bob cut the raw key into $k$ substrings of length $N$, and add these $k$ strings bitwise, obtaining the final key $K$ with length $N$. This process is the same as that in J-protocol (please see Fig.1 in Ref. \cite{QPQ11}). Till now, Bob knows the whole key $K$ and Alice generally knows only several bits in it. But if Alice is left with no known bit after this step, the protocol has to be restarted. As we will show later, if suitable parameters are selected this event happens only with small probability.

(7) Bob encrypts his database and Alice obtains the item she wanted with one of her known bits in $K$. In particular, suppose Alice knows the $j$th bit $K_j$ and wants the $i$th item (bit) of the database $X_i$. She declares the number $s=j-i$. Then Bob shifts $K$ by $s$ and using the obtained key $K'$ to encrypt his database in the manner of one-time pad. Thus $X_i$ is encrypted by $K_j$ and consequently can be correctly obtained by Alice when she gets the encrypted database.

\subsection{The features of our protocol}
It is not difficult to see that our protocol is actually a generalization of J-protocol. When $\theta=\pi/4$ our protocol becomes J-protocol, where the carrier states are $\{|0\rangle,|1\rangle,|+\rangle=\frac{1}{\sqrt{2}}(|0\rangle+|1\rangle),|-\rangle=\frac{1}{\sqrt{2}}(|0\rangle-|1\rangle)\}$ \cite{fn}. Therefore, the features of J-protocol are still hold in ours. Furthermore, because $\theta$ can be selected continuously in $(0,\pi/2)$ our protocol is more flexible and even exhibits advantages in communication complexity and security. In the following, by making the comparison to J-protocol, we discuss the features of our protocol.

On the one hand, our protocol has the same features as J-protocol in the following aspects.

(1) Different from BB84-based QPQ \cite{QPQ11}, our protocol can stand against the quantum memory attack by Alice. If Alice stores a received photon and measures it after Bob's declaration in step 4, she cannot obtain this bit because she still has to discriminate nonorthogonal states.

(2) Our protocol is loss tolerant. As discussed above, Alice has no need to tell a lie on whether one photon is lost (especially saying that a photon she has received disappeared) because before Bob's declaration she cannot get any information about the corresponding bit. Note that after Bob's declaration Alice's measurement is just like that in B92 protocol \cite{B92}. We can also directly use B92 protocol to perform QPQ (i.e. the carrier qubit is randomly in two states, $|0\rangle$ or $|0'\rangle$), where Bob's declaration is not necessary anymore. But in this condition Alice will obtain some information about the sent bit after her measurement. To elicit more bits in the raw key, for example, Alice lies that the photon on which she gets a inconclusive result is lost. So, the B92-based QPQ is not so robust against channel-loss attack.

(3) Our protocol is practical and can be easily generalized to huge-database condition. This is because Alice and Bob just execute a simple QKD protocol and no high-dimension oracle operation \cite{QPQ08,QPIR11} is needed.

On the other hand, our protocol has some peculiar merits. The primary one is its flexibility.

As in J-protocol, after adding the substrings in step 6, Alice will on average know $\bar{n}=Np^k$ ($p=\sin^2\theta/2$) bits of the final key $K$, where the number $n$ follows approximately a Poisson distribution. And the probability that she does not know any bits at all and that the protocol must be restarted is $P_0=(1-p^k)^N$. Now let us see what will happens when $\theta<\pi/4$. Without loss of generality, suppose $p=\sin^2\theta/2=0.15$. In this condition we can ensure both $\bar{n}\ll N$ and small $P_0$, which implies a successful execution of QPQ \cite{QPQ11}, by choosing an appropriate value of $k$ (see Table~\ref{tab:table1}). It can be seen that fewer substrings (i.e. a smaller $k$) are needed than that in J-protocol, which means the number of transmitted qubits are reduced. For example, when $N$=50000 only 5 substrings are needed, while in J-protocol $k$ is 7 to achieve similar $\bar{n}$ and $P_0$. Thus at least $2N=10^5$ qubits (not including the lost ones in the channel) are saved in our protocol. In fact, in our protocol $k=1$ is even feasible for not so large database size $N$ (see Tables~\ref{tab:table2} and \ref{tab:table3}). This is obviously a significant improvement with the comparison to J-protocol. Note that the condition $k=1$ will not compromise the cheat-sensitive character of the QPQ protocol, which we will discuss in the next section.

\begin{table}[b]
\caption{\label{tab:table1}%
Example of possible choice of $k$ and the values $P_0$ and $\bar{n}$ for different database sizes $N$ ($p=0.15$).}
\begin{ruledtabular}
\begin{tabular}{lllllll}
$N$ & $10^3$ & $5\times10^3$ & $10^4$ & $5\times10^4$ & $10^5$ & $10^6$\\
\colrule
$k$ & 3 & 4 & 4 & 5 & 5 & 6\\
$\bar{n}$ & 3.38 & 2.53 & 5.06 & 3.79 & 7.59 & 11.39\\
$P_0$ & 0.034 & 0.080 & 0.006 & 0.022 & $5\times10^{-4}$ & $10^{-5}$\\
\end{tabular}
\end{ruledtabular}
\end{table}

Furthermore, in our protocol the average number of the key bits Alice obtains $\bar{n}$ can be located on any fixed value the users wanted for any database size $N$. Let us recall the results in J-protocol first (see Table I in Ref. \cite{QPQ11}, which is similar to Table~\ref{tab:table1} here). When $N=50000$ we have almost no choice but letting $k=7$ and $\bar{n}=3.05$ in J-protocol. This is because $\bar{n}$ will be 12.21 if $k=6$ and 0.76 ($P_0=0.466$) if $k=8$, which are not so suitable for the aim of QPQ. The similar result exists for other $N$. But in our protocol the condition will be changed. By selecting different values of $\theta$ we can set $\bar{n}$ equals an expected value for any $N$. Tables \ref{tab:table2} and \ref{tab:table3} are the results with $k=1$ when the wanted $\bar{n}$ is 3 and 5, respectively.

\begin{table}[b]
\caption{\label{tab:table2}%
For different $N$, $\theta$ can be selected so that $k=1$ and $\bar{n}=3$.}
\begin{ruledtabular}
\begin{tabular}{llllllll}
$N$ & 12 & 50 & 100 & 200 & 500 & 1000 & 5000\\
\colrule
$p$ & 0.25 & 0.06 & 0.03 & 0.015 & 0.006 & 0.003 & 6$\times10^{-4}$\\
$P_0$ & 0.032 & 0.045 & 0.048 & 0.049 & 0.049 & 0.050 & 0.050\\
$\theta$ & 0.785($\frac{\pi}{4}$) & 0.354 & 0.247 & 0.174 & 0.110 & 0.078 & 0.035\\
\end{tabular}
\end{ruledtabular}
\end{table}

\begin{table}[b]
\caption{\label{tab:table3}%
For different $N$, $\theta$ can be selected so that $k=1$ and $\bar{n}=5$.}
\begin{ruledtabular}
\begin{tabular}{llllllll}
$N$ & 20 & 50 & 100 & 200 & 500 & 1000 & 5000\\
\colrule
$p$ & 0.25 & 0.1 & 0.05 & 0.025 & 0.01 & 0.005 & 0.001\\
$P_0$ & 0.003 & 0.005 & 0.005 & 0.006 & 0.006 & 0.006 & 0.006\\
$\theta$ & 0.785($\frac{\pi}{4}$) & 0.464 & 0.322 & 0.226 & 0.142 & 0.100 & 0.045\\
\end{tabular}
\end{ruledtabular}
\end{table}

It can be see that if we pursue $k=1$, which means the optimal communication complexity in our protocol, $\theta$ will be very small for large $N$. This might make its realization technically difficult. Therefore, $k>1$ is needed when $N$ is large. In this condition we can also locate $\bar{n}$ on an expected value for any $N$ by selecting suitable $\theta$ and $k$. For example, even though $\theta>0.2$ is required we can also achieve $\bar{n}=3$ for any $N$ by simply adjusting $k$ (see Table \ref{tab:table4}).

\begin{table}[b]
\caption{\label{tab:table4}%
For different $N$, $\theta$ $(>0.2)$ can be selected so that $\bar{n}=3$ and $P_0\approx0.05$.}
\begin{ruledtabular}
\begin{tabular}{lllllll}
$N$ & $10^3$ & $5\times10^3$ & $10^4$ & $5\times10^4$ & $10^5$ & $10^6$\\
\colrule
$k$ & 2 & 2 & 3 & 3 & 3 & 4\\
$\theta$ & 0.337 & 0.223 & 0.375 & 0.284 & 0.252 & 0.293\\
\end{tabular}
\end{ruledtabular}
\end{table}

\begin{figure}
\includegraphics[height=1.8in]{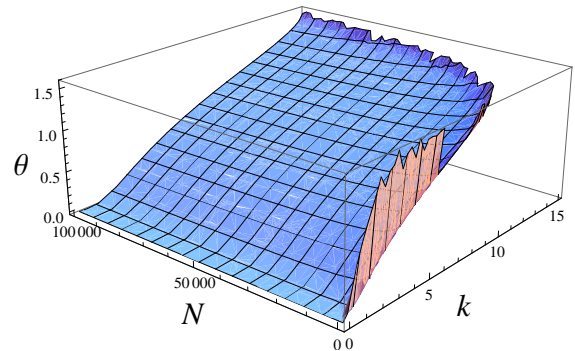}
\caption{\label{fig:one} How to achieve $\bar{n}=3$ for any $N$. Similar result also exists when we pursue $\bar{n}$ other than 3.}
\end{figure}

Fig. \ref{fig:one}, where we locate $\bar{n}=3$, is a general illustration about the flexibility of our protocol. It is shown that, we can achieve $\bar{n}=3$ by simply selecting a relatively small $\theta$, as long as it is feasible in the realization, and a suitable $k$ for any $N$.

\begin{figure}
\includegraphics[height=2in]{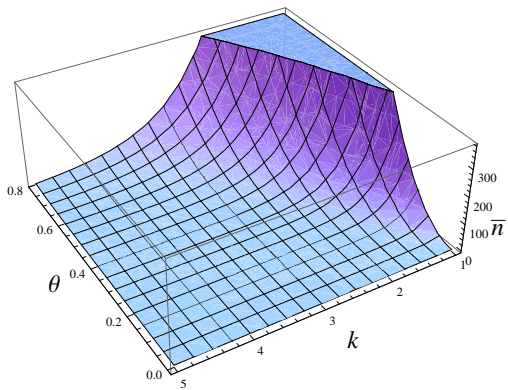}
\caption{\label{fig:two} Any $\bar{n}\ll N$ can be achieved for $N=10000$.}
\end{figure}

In fact, except for the above example for $\bar{n}=3$, $\bar{n}$ can be set on any expected number smaller than $N$ in our protocol. Sometimes we may want that Alice obtains key bits a litter more. On the one hand, as pointed in Ref. \cite{QPQ11}, Alice can use some key bits to get some other items in the database and then use them to detect Bob's potential attack. On the other hand, two users can also compare some key bits publicly (as in BB84) to check the error rate in the key. Obviously in a practical QPQ protocol Alice's final key bits may be different from Bob's, which is caused by an outside eavesdropper's attack or channel noise. Now we still have no effective way to perform error correction or privacy amplification to achieve high correctness of the final key in such a special QKD protocol (that is, Alice only gets parts of the whole key and Bob does not know which bits are obtained by Alice). Though the above comparison cannot ensure Alice's key bit, corresponding to the item she wants, is determinately the same as Bob's one, the error rate still implies the correctness of the shared key bit to some extent. For example, if the error rate is 1/10 Alice knows the key bit, which will be used to get the wanted item in the database, equals to Bob's with a high probability. On the contrary, if the error rate is higher than a threshold she will discard this result. Fig. \ref{fig:two} shows that in our protocol different $\bar{n}$ can be achieved for a fixed $N$ by adjusting $\theta$ and $k$.

Furthermore, our protocol also saves classical communication. In our protocol Bob only needs to send 1 bit, i.e. 0 or 1, to Alice for a qubit in step 4, while 2 bits are needed in J-protocol. In addition, our protocol exhibits some advantages in security, which will be demonstrated in the next section.

\section{Security analysis}

Now we consider the security of our protocol. Because it is actually a generalization of J-protocol, where $\theta=\pi/4$, the analysis in Ref. \cite{QPQ11} can be adapted here directly. It is not difficult to imagine that our protocol can also ensure the security of the database and the privacy of the user. In the following we focus on how the security degree changes when $\theta\neq\pi/4$ in our protocol.

\subsection{Database security}
If Alice is dishonest and she wants to obtain more items in Bob's database, she has to try to obtain more key bits in the raw key $K^r$. To this aim Alice can store the qubits received from Bob and take more effective measurements on them after Bob's declaration in step 4.

\begin{figure}
\includegraphics[height=1.5in]{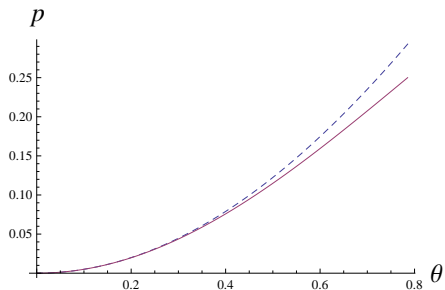}
\caption{\label{fig:three} Comparison between Alice's success probability by USD measurement and projective one on a qubit. The dashed line represents the result for USD measurement and the solid line is for projective one.}
\end{figure}

Let us consider the simple measurement for Alice first, i.e. individual one for each qubit she received. For example, if Bob said a qubit is in one of the states $\{|0\rangle, |0'\rangle\}$ by declaring 0, Alice performs the optimal unambiguous state discrimination (USD) measurement \cite{USD1,USD2} to distinguish which state the qubit is in. The success probability of this USD measurement is bounded by $1-F(\rho_0,\rho_1)$, where $F(\rho_0,\rho_1)$ is the fidelity between the two states to be discriminated. So in our protocol this probability is $p^{USD}=1-\langle0|0'\rangle=1-\cos\theta$. It can be seen that, the advantage Alice obtains by USD measurement is negligible compared with the legal projective one, where the probability is $p=\sin^2\theta/2$, especially when $\theta$ is small (see Fig. \ref{fig:three}). For example, when $\theta=0.284$, $N=50000$ and $k=3$ (see Table \ref{tab:table4}), Alice can get $\bar{n}^{USD}=50000\times(1-\cos0.284)^3=3.21$ bits of the key by USD measurement, which is just a little more than the value by projective one, i.e. $\bar{n}=50000\times[(\sin0.284)^2/2]^3=3.02$. Note that this implies an improvement compared to J-protocol, where $\bar{n}^{USD}=9.3$ in the same situation, i.e. pursuing $\bar{n}=3$.

\begin{figure}
\includegraphics[height=1.5in]{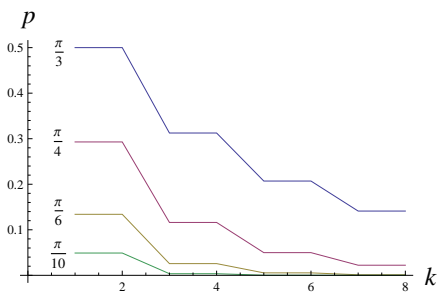}
\caption{\label{fig:four} Alice's success probabilities to obtain the final key bits by joint USD measurement for different $\theta$. Among them the line denoted as $\theta=\pi/4$ is the result of J-protocol. It can be seen that when $\theta$ is small, Alice will get less bits in the final key, which implies a higher security degree for the database under this kind of attack.}
\end{figure}

As analyzed in Ref. \cite{QPQ11}, Alice can also perform joint measurement on the $k$ qubits which contribute to an element of the final key. By this means she wants to obtain the bit value of the final key directly without distinguishing the individual bit values of the raw key. In this condition two kinds of measurements can be used by Alice. One is Helstrom's minimal error-probability measurement, i.e., the measurement that distinguishes two quantum states with the highest information gain \cite{MM1,MM2}. To distinguish two equally likely quantum states $\rho_0$ and $\rho_1$, the
probability to guess the state correctly is bounded by $p_{guess}=\frac{1}{2}+\frac{1}{2}D(\rho_0, \rho_1)$, where $D(\rho_0, \rho_1)$ is the trace distance between $\rho_0$ and $\rho_1$. In our protocol, therefore, Alice can correctly guess a final key bit with the probability at most $p_{guess}=\frac{1}{2}+\frac{1}{2}\sin^k\theta$. Obviously when $\theta$ is small this probability is close to 1/2 (a random guess). The other measurement Alice can use is USD measurement. In this condition the success probability of unambiguously discriminating the two $k$-qubit mixed states corresponding to odd and even parity can be obtained, which declines rapidly with $k$ (see Fig. \ref{fig:four}). In Fig. \ref{fig:four} the probabilities that Alice can get the final key bits by joint USD are depicted for different values of $\theta$, where we can see that when $\theta$ is small Alice's advantage by joint USD is distinctly decreased.

To sum up, if $\theta<\pi/4$ is chosen in our protocol it exhibits an obvious improvement compared to J-protocol in terms of the database security.

\subsection{User privacy}

In QPQ protocols, the user privacy is pursued in the manner of cheat-sensitive \cite{QPQ08,QPIR11,QPQ11}. That is, Bob will run the risk of being discovered if he tries to obtain the address queried by Alice. In GLM protocol such a dishonest Bob will inevitably send unsatisfied answer to Alice and then be discovered with a certain probability. While in O-protocol and J-protocol dishonest Bob might send wrong answer to Alice, which will also be detected by Alice at a later time.

Our protocol is also cheat sensitive for user privacy in the sense that Bob cannot simultaneously obtain the query address and give right answer for the query deterministically. The reason is the same as that in Ref.\cite{QPQ11}. If dishonest Bob, by sending fake states or performing special measurements, can get both the conclusiveness of one of Alice's elements of the raw key and the corresponding value of this key bit, he knows the exact measurement basis chosen by Alice. Thus, Bob knows Alice's choice once she measured the qubit randomly in one of two bases, which implies superluminal communication. Therefore, the security of user privacy in our protocol is assured by the no-signaling principle.

\begin{figure}
\includegraphics[height=1.5in]{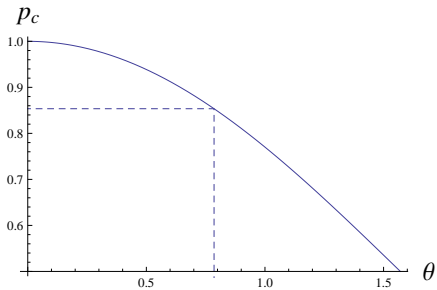}
\caption{\label{fig:five} The relation between the probability $p_c$, with which Alice gets conclusive result, and the value of $\theta$ under Bob's attack. The dashed line denotes the result in J-protocol (i.e. $\theta=\pi/4$).}
\end{figure}

In the following we briefly discuss the probability with which Bob can obtain the conclusiveness of any of Alice's bits in the raw key, and demonstrate the relation between this probability and the value of $\theta$. Similar to the analysis in Ref. \cite{QPQ11}, Bob can get the conclusiveness of Alice's one bit with the optimal probability by sending $|0''\rangle$ ($|1''\rangle$) and announcing 1 (0) in step 4. Here
\begin{eqnarray}
|0''\rangle=\cos(\theta/2)|0\rangle+\sin(\theta/2)|1\rangle, \nonumber\\
|1''\rangle=\sin(\theta/2)|0\rangle-\cos(\theta/2)|1\rangle.
\label{eq:two}
\end{eqnarray}
Thus Bob knows that Alice will get conclusive result on this qubit with the probability $p_c=\cos^2(\theta/2)$. If Bob expects that Alice gets inconclusive result on one qubit, he just sends $|0''\rangle$ ($|1''\rangle$) and announcing 0 (1) in step 4. By this method the probability with which Alice gets conclusive result equals $p_i=1-p_c=\sin^2(\theta/2)$. Fig. \ref{fig:five} demonstrates the relation between $p_c$ and $\theta$. It can be seen that a smaller $\theta$ implies a higher probability with which Bob can predict Alice's conclusive bits. As analyzed above, we are inclined to use a small $\theta\in(0,\pi/4)$ to achieve its advantages in communication complexity and security degree of the database. In this condition Bob generally can obtain the address of Alice's query with relatively higher probability. But this does not decrease Alice's privacy in the sense that it is still cheat sensitive. This is because Bob will inevitably lose the knowledge about the value of the key bit when he tries to obtain its conclusiveness, which is the same as that in J-protocol. For example, in the above attack, Bob will completely not know the value of the corresponding key bit though he can optimally estimate the address of Alice's query. Consequently it is impossible for Bob to give Alice a right answer with certainty. This is assured by the no-signaling principle.

Generally speaking, as pointed out in Sec.II, the users can choose a suitable value of $\theta$ so that $k=1$ in our protocol, which means the optimal communication complexity. Recalling the process in step 6, we know that it is easier for Bob to guess the conclusiveness of one bit in Alice's final key when $k$ is smaller. But this fact does not hurt the security of this protocol when $k=1$ because Bob will lose the value of the bit when he tries to get its conclusiveness. Of course, a large $\theta$ (e.g. $\theta>\pi/4$) can also be selected to achieve a small $p_c$ if we pursue a low probability with which Bob can correctly guess the address of Alice's query (see Fig. \ref{fig:five}). This exhibits the flexibility of our protocol again.

\section{Conclusions}

We present a flexible QKD-based QPQ protocol, which can be seen as a generalization of J-protocol \cite{QPQ11}. It retains all the features of J-protocol. For example, it is loss tolerant, practical and robust against quantum memory attack. Furthermore, our protocol is very flexible and controllable. When a small $\theta$ ($\theta<\pi/4$) is used the aim of QPQ can be achieved with a smaller $k$ (even $k=1$) than that in J-protocol, which implies lower complexity of both quantum and classical communications. In our protocol, by adjusting the value of $\theta$, the average number of the key bits Alice obtains can be located on any fixed value the users wanted for any database size $N$. In addition, when $\theta<\pi/4$ our protocol exhibits better database security, which is ensured by the impossibility of perfectly distinguishing
nonorthogonal quantum states. At the same time, user privacy is ensured by the no-signaling principle and, in some special conditions, $\theta>\pi/4$ can be also selected to obtain a low probability with which Bob can correctly guess the address of Alice's query.

\section*{Acknowledgements}
This work is supported by NSFC (Grant Nos. 61170270, 61100203, 60903152, 61003286, 60821001), NCET (Grant No. NCET-10-0260), SRFDP (Grant No. 20090005110010), Beijing Natural Science Foundation (Grant No. 4112040), the Fundamental Research Funds for the Central Universities (Grant Nos. BUPT2011YB01, BUPT2011RC0505, 2011PTB-00-29), Science and Technology on Communication Security Laboratory Foundation (Grant No. 9140C110101110C1104).

\end{document}